\documentclass[letter,traditabstract]{aa} % for the letters
\usepackage{graphicx}
\usepackage{txfonts}
\usepackage{natbib}
\bibpunct{(}{)}{;}{a}{}{,}
\usepackage{comment}
\begin{document}
   \title{Constraints on the optical precursor to the naked-eye burst GRB080319B from ``Pi~of~the~Sky'' observations}

   \author{Lech Wiktor Piotrowski}

   \institute{Faculty of Physics, University of Warsaw,
	Ho\.za 69, 00-681 Warsaw, Poland
             }
 
\authorrunning{L. W. Piotrowski}
\titlerunning{Constraints on the optical precursor to the naked-eye burst}

\date{Received: 15 December 2011 / Accepted: 7 February 2012}

\abstract
{
	I present the results of the search for an optical precursor to the naked-eye burst -- GRB080319B, which reached $5.87^{\rm m}$ optical peak luminosity in the ``Pi of the Sky'' data. A burst of such a high brightness could have been preceded by an optical precursor luminous enough to be in detection range of our experiment. The ``Pi of the Sky'' cameras observed the coordinates of the GRB for about 20 minutes prior to the explosion, thus provided crucial data for the precursor search. No signal within $3\sigma$ limit was found. A limit of $12^{\rm m}$ (V-band equivalent) was set based on the data combined from two cameras, the most robust limit to my knowledge for this precursor.
}
%{no}{no}{no}{no}

   \maketitle
%
%________________________________________________________________

\section{Introduction}

Observations of the optical emission accompanying gamma-ray bursts have played a crucial role in understanding this phenomenon. The measurement of the distance to these explosions has allowed us to estimate their true energy and thus significantly constrain the models describing their origin \citep{first_host}. The presence of very bright  optical counterparts during the gamma-ray emission was first established by \cite{aft_cor} and the existence of  prompt optical emission associated with the prompt gamma-ray was established in two papers by \cite{first_sim,second_sim}. Optical observations of the naked-eye burst GRB080319B shed new light on the prompt emission mechanism, revealing for the first time, that a very bright optical counterpart became detectable simultaneously (with $\pm 5$ s precision) with the $\gamma$-ray emission \citep{nature}. However, much remains unclear about this phenomenon including the nature of the central engine or interpretation of some of the observed features, such as $\gamma$-ray precursors visible for about $20\%$ of GRBs \citep{batse_prec}. These precursors, as well as prompt emission, may coexist with optical precursors.

The aforementioned precursors have not yet been observed and had hardly any chance to be seen due to minimal amount of optical observations preceding the GRB explosion. However, there are some models predicting the existence of a precursor. According to these models, precursor detections could explain some emission features not yet understood \citep{spinar} or, more importantly, reveal crucial information about the nature of the GRB central engine \citep{QN2} and answer some long-standing questions. Therefore, the discovery of an optical precursor could revolutionise our knowledge of GRBs.

Investigations to date of possible optical precursor mechanisms have reached no firm conclusion. However, in my opinion, two main types of optical precursors models can be distinguished. The first attributes the precursor generation to the activity of the central engine, such as that produced by a two-stage collapse. In one group of these models, the burst is preceded by a progenitor star ending its life in a supernova Ibc explosion \citep{supernova_con}. The explosion should be visible as an optical precursor, although its luminosity may be similar to those of standard supernovae and too low, owing to the distance of most of bursts observed by optical observatories \citep{supernova_num}. The time between the precursor and the main burst depends strongly on the scenario of the explosion, in the case of the SupraNova model \citep{SupraNova} even being as long as months to years, but some argue that it can be much shorter \citep{supranova_times}. The SupraNova models also allow for optical precursors caused by plerionic activity, which can reach $17^{\rm m}$ for $z\sim1$ in R-band and exist for very short times before the GRB \citep{plerion}.

Somewhat similar groups of models predict a supernova explosion and the creation of a neutron star, which then undergoes a ``Quark-nova'' explosion leading to a GRB \citep{QN}. These models allow for precursor emission seconds before the main outburst \citep{QN2} in optical, X-ray, and $\gamma$-ray regimes and the observation of an optical precursor could yield crucial information on the delay between the SN and QN. A different model \citep{spinar}, still concentrating on the activity of the central engine, analyses a collapse of a so-called ``spinar'' -- a rotating object for which equilibrium is the result of an interaction between centrifugal and gravitational forces. The model also permits optical precursors to appear during the lifetimes of observed $\gamma$-ray precursors.

The other branch of models predicting an optical precursor focuses on the analysis of the outflow itself, not the central engine. Some models, such as that of \cite{thermal}, predict $\gamma$-ray precursors a with thermal spectrum. However, the precursor itself would be 10-100 times less intense than the main burst, so its optical tail would be very weak. On the other hand, \cite{nonthermal} found that there may be a non-thermal optical precursor, which may however be very weak.

The detection of GRB080319B enhanced our belief in an observable optical precursor, as for the first time this object provided the astronomical community with an indisputable proof that the initial $\gamma$ emission of the gamma-ray bursts can be accompanied by strong optical emission. The burst was moderate in terms of gamma radiation, but extremely luminous, reaching $5.3^{\rm m}-5.87^{\rm m}$ (depending on the experiment) at visible wavelengths. When related to the distance $z=0.937$, the luminosity is the highest observed to date, beating the previous record set by SN 2005ap by a factor of $2.5\cdot10^6$. The burst was also exceptionally bright at X-rays. Nevertheless, an extrapolation of $\gamma$-ray spectrum to optical energies leads to fluxes much less intense than the observed optical radiation, which rules out the possibility, that the optical flash is the low energetic tail of the high energy emission. This observation together with a possible correlation between peaks in both bands provides support to calls for a revision of prompt emission models \citep{nature, tortora_var}.

The predictions of optical precursor models are limited in scope, most likely owing to the very few observations providing any limits on this type of emission, such as \cite{precursors}. Perhaps the main motivation for the search for this precursor is the knowledge acquired from studies of the naked-eye burst that the optical emission can be simultaneous with the $\gamma$-ray emission. If this is due to the coexistence of emission phenomena in both bands \citep{simult}, there is a possibility, that the observed $\gamma$-ray precursors are also accompanied by optical precursors. In most cases, the observed optical emission was less luminous than the $\gamma$-ray emission and, the optical precursor, if accompanying $\gamma$-ray would be too weak to be detected. However, the extreme brightness of GRB080319B in the optical band, relative to the moderate intensity in $\gamma$ band hints that it could be possible to have an optical precursor bright enough to be seen, even if the $\gamma$-ray precursor was not visible to experiments.

Only two constrains on the optical precursor emission to GRB080319B have so far been published. The coordinates of the burst were observed for 40 minutes starting more than 2 hours before the detection by 152 cm telescope located in Loiano \citep{limit1}. The limiting magnitude of four coadded 10 minute exposures in filter R is $20.3^{\rm m}$. However, this observation does not exclude the existence of much brighter, but very short precursors. The all-sky monitor RAPTOR-Q has set brightness limits in the time period covering known $\gamma$-ray precursor times -- 30 minutes prior to the main outburst -- to 0.5 Jy ($\sim 9.63^{\rm m}$ in V filter). The TORTORA experiment set limits of $8.5^{\rm m}$ for 100 seconds prior to the SWIFT trigger \citep{tortora_limit}. The presented results improve these limits significantly.

In all the computations in this article, the $\Lambda_{CDM}$ model has been used, with $H_0=71$, $\Omega_m=0.27$, and $\Omega_{\Lambda}=0.73$.

\section{The ``Pi of the Sky'' project}

While the lack of discovery of an optical precursor to date may be attributed to the nature of a GRB itself, the poor constrains on the phenomenon occurring between second and dozens of minutes before GRB is a consequence of present optical observation strategies. Nearly all of the optical GRB observatories are follow-up experiments. Detectors rotate to the coordinates of the burst after receiving a signal from a GRB orbital experiment, such as SWIFT BAT or Fermi GBM. This strategy in addition to the small field of view of these experiments make observation of the optical prompt emission difficult, being possible only for very few long bursts and with very fast devices, and nearly rule out the possibility of observing an optical precursor. This is one of the main reasons why ``Pi of the Sky'' project adopts a completely different strategy.

The ``Pi of the Sky'' experiment is designed for continuous monitoring of a large fraction of the sky with high time resolution ($\sim10$ s). A real-time analysis of the data stream, based on a multi-level triggering system, allows discoveries of GRB optical counterparts independently of satellite experiments \citep{pi1,pi2}. This approach resulted in the autonomous detection of the naked-eye burst GRB080319B at its very beginning \citep{nature}. The strategy also allows the search for optical precursors to GRBs, similar to those described in this article.

The GRB080319B was detected by the prototype system during its first work period (2004-2009) in Las Campanas Observatory in Chile. The prototype consists of two cameras placed on a paralactic mount, observing the same part of the sky to allow the elimination of optical flashes due to cosmic radiation hitting the CCD sensor.

To meet the requirement for monitoring a large fraction of the sky, the ``Pi of the Sky'' apparatus makes use of cameras with a very wide field of view -- about $20^\circ \times 20^\circ$ each. For stars positioned far from the optical axis, this causes significant deformations of images, which are much larger than in other astronomical experiments. This was also the case for GRB080319B, for which the position of the burst was in the corner of the frame up to $t_0+36$ s. The observed precursor  would therefore also be deformed and thus large uncertainties would be introduced into standard photometric and signal-searching algorithms.

\begin{figure}
   \resizebox{0.49\hsize}{!}{\includegraphics{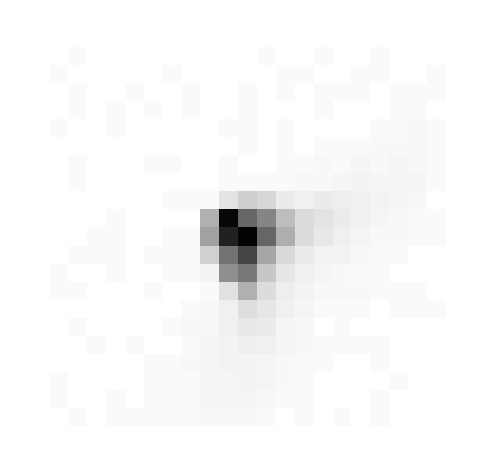}}
	\resizebox{0.49\hsize}{!}{\includegraphics{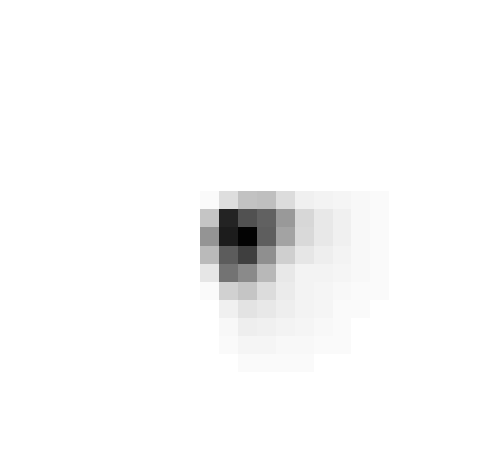}}
      \caption{Comparison of the real (left) and simulated (right) GRB080319B images (in peak brightness), the second of which was close to the frame corner, where the PSF is most deformed. The simulated PSF, used for the precursor search described in this paper, reproduces the real PSF in great detail.}
         \label{simul_real}
   \end{figure}

To improve the brightness and coordinate measurements, a model of the PSF in the ``Pi of the Sky'' system was created. A standard method used to determine the PSF shape is based on the superposition
of real sky star images. However, in the case of very deformed stars, it introduces large uncertainties because the exact shape is unknown and the star position is thus uncertain.

Our approach was therefore different. First, the PSF profiles of a point-like source were precisely measured across the CCD chip in laboratory conditions and reconstructed with high resolution. Second, a model based on Zernike polynomials was fitted to high-resolution profiles \citep{phd}. Finally, the laboratory-based model parameters were refitted to real sky images, to accurately reproduce the deformations that are characteristic of a given camera.

   \begin{figure}
%   \centering
   \resizebox{\hsize}{!}{\includegraphics{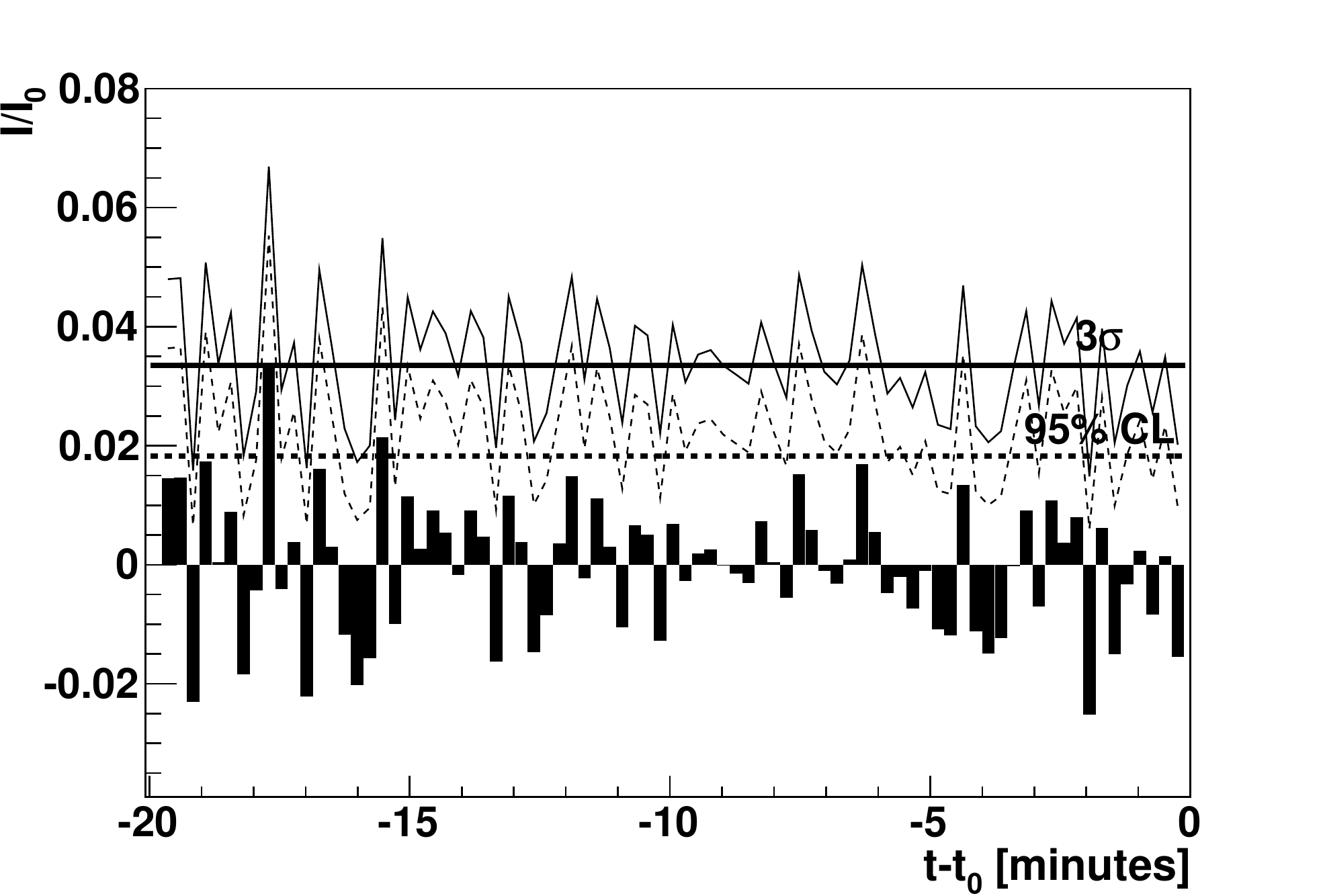}}
      \caption{Signal value $\rm \frac{I}{I_0}$ at the GRB080319B position relative to the nearby reference star, obtained from the PSF profile fit to the data from the k2a camera of the ``Pi of the Sky'' prototype as a function of time before the burst $\rm t-t_0$ (vertical bars). We also indicate the limits on the precursor luminosity calculated assuming no signal (thick lines) and taking the actual signal measured for each frame (thin lines). Limits are calculated at both the $3\sigma$ level (solid lines) and $95\%$ confidence level (dashed lines). }
         \label{signal_k2a}
   \end{figure}

The final model can be used to improve photometric and astrometric measurements, and fitted to the given coordinates to perform a detailed search for a signal. Additionally, the model was used to create a sophisticated frame simulator, namely one able to reproduce the frame with the expected star PSFs, several types of noises and fluctuations, etc. The simulator can be used to verify analysis methods and design future hardware \citep{karlove}.

Simulated PSFs obtained in this way are very close to real star images even                                                                                     for the most deformed stars, as can be seen in fig. \ref{simul_real}. As the model was tuned to the limited pixel range around the star centre, it does not fully account for the long tails of the PSF or a halo appearing around very bright stars. However, these effects have a negligible influence on the position and brightness measurements, as a vast majority of the signal is measured in the accurately modelled part of PSF.

\section{Search for the GRB080319B optical precursor}

The search was performed by fitting the model PSF at GRB coordinates to all the frames covering 19 minutes prior to the explosion, on two cameras (with the internal names k2a and k2d) of the ``Pi of the Sky'' prototype. Figure \ref{signal_k2a} shows the fitted signal value of $\rm \frac{I}{I_0}$ for the k2a camera, for all considered frames. To suppress the systematic uncertainty, the fitted signal I is referred to the scale $\rm I_0$ of a nearby $7.99^{\rm m}$ star (HIP 70912 in Hipparcos Catalogue \citep{Hipparcos}), given in the VT (visual Tycho) filter\footnote{The ``Pi of the Sky'' prototype cameras were not equipped in any filter, except for an IR+UV cut filter. This resulted in a relatively wide spectral sensitivity of the detector, with average $\lambda\simeq586$ nm and a spread of about $77$ nm, closest to the VT and Johnson V filters \citep{pi_filter}.}. No signal exceeding the $3\sigma $ limit has been found for this camera. The standard approach in the case of ``no signal'', at least in astronomical observations, is to quote a $3\sigma$ limit assuming a measured value of signal to be zero. The estimation of all the limits is based on an assumption that the measurement error for a small signal that would be emitted by the optical precursor is similar to the fit error for the sky background at this position. The uncertainty $\sigma$ was calculated from a fit of the normal distribution to the histogram of signal ratio $\rm \frac{I}{I_0}$. The result is consistent with distributions of signal in three empty control areas in the burst proximity that were considered as a cross-check. In this approach, the resulting $3\sigma$ limiting magnitude for the k2a camera in the polynomial photometry is $11.67^{\rm m}$. At the $95\%$ confidence level ($=1.96\sigma$), the limiting magnitude for k2a camera is $12.13^{\rm m}$.

   \begin{figure}
%   \centering
   \resizebox{\hsize}{!}{\includegraphics{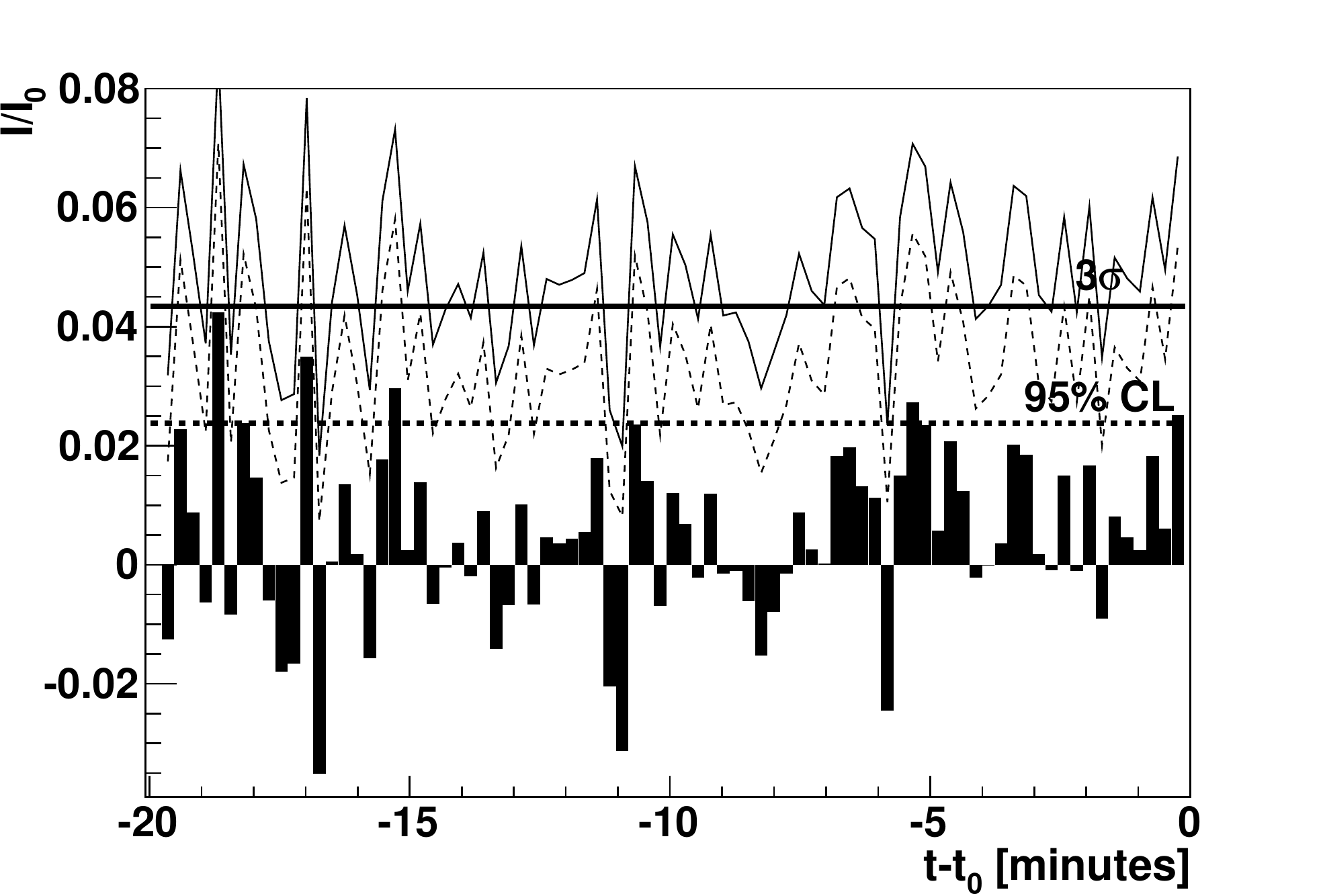}}
      \caption{As in Fig. \ref{signal_k2a} but for the k2d camera of the ``Pi of the Sky'' prototype.}
         \label{signal_k2d}
   \end{figure}

Figure \ref{signal_k2d} shows the fitted signal value $\rm \frac{I}{I_0}$ for the k2d camera, for all considered frames. The performance of this camera was a little bit worse than the k2a, so the background fluctuations are higher. Thus, the limits are slightly worse, the resulting $3\sigma$ limiting magnitude, assuming no signal, being $11.39^{\rm m}$. In this approach, any source brighter then $11.85^{\rm m}$ is excluded at the $95\%$ confidence level (CL).

The method that we used also allowed us to set limits on particular frames based on the fitted signal level in the GRB coordinates (the fit results changing from frame to frame owing to the background fluctuation where we seek the signal). The limit was numerically calculated according to the formula given in \cite{stat_limits}, which allows us to extract limits with a well-defined confidence level (CL) for any value of the measured signal, including negative signal fluctuations. The $3\sigma$ limits ($99.73\%$ CL) calculated on single frames of the k2a camera fluctuate in most cases between $11.5^{\rm m}$ and $12.25^{\rm m}$. The $95\%$ confidence level limits (which seem more appropriate for a single-frame analysis) fluctuate in most cases between $11.6^{\rm m}$ and $13^{\rm m}$. For the k2d camera, the corresponding limits fluctuate between $11^{\rm m}$ and $12.3^{\rm m}$ ($3\sigma$ limits) and $11.2^{\rm m}$ and $13.1^{\rm m}$ ($95\%$ confidence level limits) in most cases.

\begin{figure}
%   \centering
   \resizebox{\hsize}{!}{\includegraphics{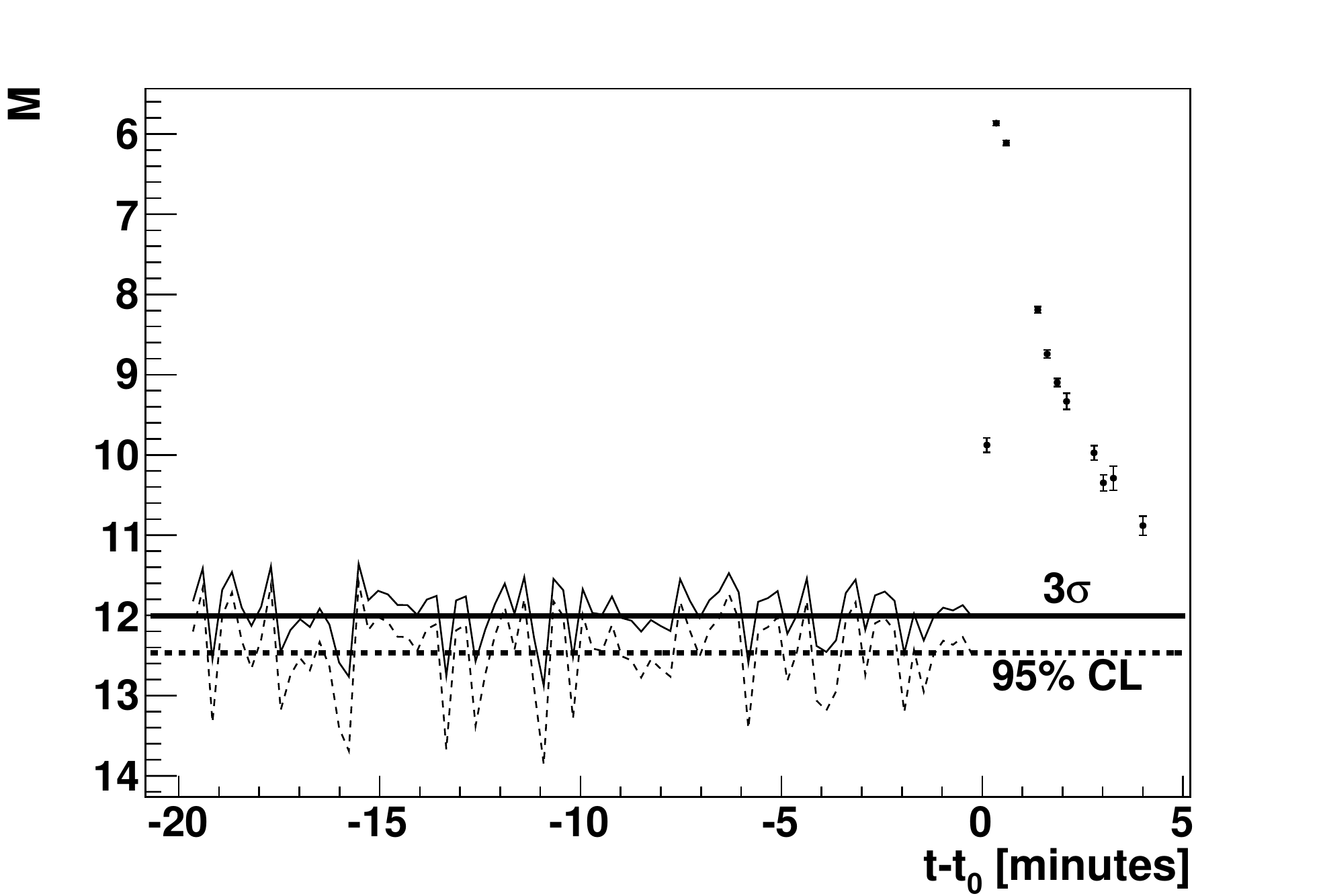}}
      \caption{Limiting magnitude M (V-band equivalent) for the optical precursor emission from GRB080319B as a function of time before the burst $\rm t - t_0$. Limits to the precursor luminosity were calculated assuming no signal (thick lines) and taking the actual signal measured for each frame (thin lines). Limits were calculated at the $3\sigma$ level (solid lines) and $95\%$ confidence level (dashed lines) and are based on combined data from two cameras of the ``Pi of the Sky'' prototype. The points with error bars represent the GRB080319B ``Pi of the Sky'' flux \citep{nature}.}
         \label{limits}
   \end{figure}

The combined scale for a precursor was then computed as the weighted average of scales fitted for both cameras. No signal above $3\sigma$ level was found in the combined signal distribution. However, most limits greatly improve, as shown in fig. \ref{limits}, where limits obtained in this analysis are compared to the measured GRB080319B lightcurve \citep{nature}. The standard $3\sigma$ limit calculated assuming a zero signal is $12^{\rm m}$, which is fainter by $0.33^{\rm m}$ than the single k2a camera limit, and by $0.61^{\rm m}$ than the single k2d camera limit. The $95\%$ CL limit also increases by $0.33^{\rm m}$ relative to the k2a limit and $0.62^{\rm m}$ relative to that of k2d, reaching $12.47^{\rm m}$. The $3\sigma$ limit based on the measured signals for single frames is between $11.5^{\rm m}$ and $12.6^{\rm m}$ for most frames. The $95\%$ CL limit for most frames is contained between $11.7^{\rm m}$ and $13.3^{\rm m}$.

In general, the combined limits based on signals measured in the two cameras are more stable than limits for the single, k2a camera. The measurement error was reduced by nearly $30\%$, from $0.011$ for k2a to $0.008$ for the combined signal. This is the main reason for the limit improvement with the two cameras.

\section{Conclusions}

The coordinates of GRB080319B -- the naked-eye burst -- were observed 19 minutes prior to the explosion. The data have been analysed with a dedicated signal search algorithm, based on a sophisticated model of the ``Pi of the Sky'' PSF. Although no signal exceeding $3\sigma$ limit has been found for either single camera or the combined measurements, new, stronger limits on precursor luminosity have been established. The combined $3\sigma$ limit, assuming no signal is a $12^{\rm m}$ apparent magnitude, being just $0.35\%$ of the peak ``Pi of the Sky'' flux, which is a significant improvement compared to the limit set by RAPTOR-Q in the similar time period, namely $2.5\%$ of the peak flux \citep{limit_raptor}. The rough estimation gives the absolute magnitude precursor limit of $-31.9^{\rm m}$.

The observation is consistent with the plerionic precursor model, which gives $~\sim17^{\rm m}$ optical precursor in our spectral range for GRB at $z\sim1$ \citep{plerion}. For the naked-eye burst ($z=0.937$), this gives a rough estimate of the $16.8^{\rm m}$ precursor luminosity (with the same cosmological assumptions as above), far beyond the ``Pi of the Sky'' range. The expected supernova explosions at the GRB080319B distance would also be beyond our range, even if the event were superluminous, as in the case of the expected quark-novae \citep{QN2}.

On the other hand, if we assume that the optical precursor luminosity is proportional to the main peak luminosity, and the proportion is similar to that observed for $\gamma$-ray emission, i. e. 10-100 times dimmer than the peak luminosity \citep{thermal}, we can rule out the possibility of GRB080319B having an optical precursor. According to our presented calculations, the precursor would have to be at least 300 times dimmer than the peak luminosity. Future calculations, which were beyond the scope of this paper, for specific models of this burst should be performed and compared to the aforementioned limit, to provide improved constrains on models parameters.

\begin{acknowledgements}
The author would like to thank the whole ``Pi of the Sky'' team, who made the observations required for this work possible. Special gratitude should be conveyed to Aleksander Filip \.Zarnecki and Agnieszka Pollo for support and sharing their knowledge during the preparation of this article. This work was supported by the Polish Ministry of Science and Higher Education in 2009-2011 as a research project.
\end{acknowledgements}

\bibliographystyle{aa}
\bibliography{paper}

\end{document}